\documentclass[12pt]{article}
\usepackage[dvips]{graphicx}

\begin{document}

 \title{ Universal property of the information entropy in
 fermionic and bosonic systems.}

\author{S.E. Massen, Ch.C. Moustakidis and C.P. Panos\\
Department of Theoretical Physics,\\
Aristotle University of Thessaloniki,\\
GR 54006 Thessaloniki, Greece}

\maketitle

\begin{abstract}

 It is shown that a similar functional form $S=a+b\ln N$ holds
 approximately for the information entropy $S$ as function of the
 number of particles $N$ for atoms, nuclei and atomic clusters
 (fermionic systems) and correlated boson-atoms in a trap (bosonic
 systems). It is also seen that rigorous inequalities previously
 found to hold between $S$ and the kinetic energy $T$ for
 fermionic systems, hold for bosonic systems as well. It is found
 that Landsberg's order parameter $\Omega$ is an increasing
 function of $N$ for the above systems. It is conjectured that the
 above properties are universal i.e. they do not depend on the
 kind of constituent particles (fermions or correlated bosons) and
 the size of the system.

 \end{abstract}

 Shannon's information entropy for a continuous probability
 distribution $p(x)$ is defined as
\begin{equation}
S= - \int p(x) \ln p(x) d x
 \label{A-shan}
\end{equation}
where $\int p(x) d x =1$. $S$ is measured in bits if the base of
the logarithm is 2 and nats (natural units of information) if the
logarithm is natural.

This quantity is useful for the study of quantum systems
\cite{Ohya93,Bialynicki75,Massen98,Massen01,Gadre85,Gadre87,Ghosh84,%
Lalazissis98,Moustakidis01,Panos01a,Panos01b} and appears in
various areas: information theory, ergodic theory and statistical
mechanics. It is closely related to the entropy and disorder in
thermodynamics. It represents the information content or
uncertainty of $p(x)$ and has already been connected with
experimental and/or fundamental quantities (e.g. the kinetic
energy and magnetic susceptibility in atomic physics
\cite{Gadre85,Gadre87} and the kinetic energy and mean square
radius in nuclear and cluster physics \cite{Massen01}).

In recent years information-theoretic methods play an increasing
role for the study of quantum mechanical systems
\cite{Ohya93,Bialynicki75,Massen98,Massen01,Gadre85,Gadre87,Ghosh84,%
Lalazissis98,Moustakidis01,Panos01a,Panos01b}.
An important step was the discovery of an entropic uncertainty
relation (EUR)\cite{Bialynicki75}, which for a three-dimensional
system has the form
\begin{equation}
S= S_r + S_k \ge 3 (1+ \ln \pi) \simeq 6.434 \quad (\hbar=1)
 \label{S1}
\end{equation}
where $S_{r}$ is the information entropy in position space of the
density distribution $\rho ({\bf r})$ of a quantum system
\begin{equation}
S_{r}= - \int \rho({\bf r}) \ln \rho({\bf r}) d {\bf r}
 \label{Sr1}
\end{equation}
and $S_{k}$ is the information entropy in momentum space of the
corresponding momentum distribution $n({\bf k})$
\begin{equation}
S_{k}= - \int n({\bf k}) \ln n({\bf k}) d {\bf k}
 \label{Sk1}
\end{equation}

The density distributions $\rho({\bf r})$ and $n({\bf k})$ are
normalized to one. It is noted that the entropy sum $S$ does not
depend on the units in measuring $r$ and $k$, i.e. it is scale
invariant to a uniform change of coordinates.
Inequality (\ref{S1}) provides a lower bound for $S_{r}+S_{k}$,
which is attained by Gaussian density distributions. Its physical
meaning is obvious: an increase of $S_{k}$ is accompanied by a
decrease of $S_{r}$ and vice versa, which indicates that a diffuse
$n({\bf k})$ is associated with a localised $\rho({\bf r})$ and
vice versa. This is also expected from the Heisenberg uncertainty
relation, but (\ref{S1}) is a strengthened version of the
uncertainty principle.

The present Letter addresses the problem of finding $S_{r}$,
$S_{k}$ (i.e. the extent of $\rho({\bf r})$ and $n({\bf k})$) for bosonic
many-body sytems in order to compare with corresponding results
for fermionic systems. First, we review previous work for systems
of fermions.

In \cite{Massen98} we proposed a universal property for $S$ for
the density distributions of nucleons in nuclei, electrons in
atoms and valence electrons in atomic clusters. This property has
the form
\begin{equation}
S=a+b \ln N
 \label{S-ln1}
\end{equation}
where the parameters $a$, $b$ depend on the system under
consideration. The values of the parameters are the following
\begin{eqnarray}
a=5.325& \quad b=0.858 & \quad ({\rm nuclei}) \nonumber\\ a=5.891&
\quad b=0.849 & \quad ({\rm atomic \ clusters})\\ a=6.257& \quad
b=1.007 & \quad ({\rm atoms})  \nonumber \label{eq-8}
\end{eqnarray}

For the total densities, there is in atomic physics a connection
of $S_{r}$, $S_{k}$ with the total kinetic energy $T$ and mean
square radius of the system through rigorous inequalities derived
using the EUR \cite{Gadre87}
\begin{eqnarray}
 \label{Sr-ineq}
 &&S_r({\rm min}) \le S_r \le S_r({\rm max})\\
 \label{Sk-ineq}
 &&S_k({\rm min}) \le S_k \le S_k({\rm max})\\
 \label{S-ineq}
 &&S({\rm min}) \le S \le S({\rm max})
\end{eqnarray}

The lower and upper limits are written here more conveniently in
the following form, for density distributions normalized to one:
\begin{eqnarray}
S_r({\rm min})&=& \frac{3}{2}(1+\ln \pi)
           -\frac{3}{2} \ln \left( \frac{4}{3} T \right) \nonumber\\
S_r({\rm max})&=& \frac{3}{2}(1+\ln \pi)
           +\frac{3}{2}\ln \left(\frac{2}{3}\langle r^2 \rangle \right)
 \label{Sr-min}
\end{eqnarray}
\begin{eqnarray}
S_k({\rm min})&=& \frac{3}{2}(1+\ln \pi)
           -\frac{3}{2}\ln \left( \frac{2}{3}\langle r ^2\rangle \right) \nonumber\\
S_k({\rm max})&=& \frac{3}{2}(1+\ln \pi)
           +\frac{3}{2}\ln \left(\frac{4}{3} T \right)
            \label{Sk-min}
\end{eqnarray}
\begin{eqnarray}
S({\rm min})&=& 3(1+\ln \pi)  \nonumber\\
S({\rm max})&=& 3(1+\ln \pi)
        + \frac{3}{2}\ln \left(\frac{8}{9}\langle r^2\rangle T \right)
 \label{S-min}
\end{eqnarray}

In a previous Letter \cite{Massen01} we verified numerically that
the above inequalities hold for nuclear density distributions and
valence electron distributions in atomic clusters. We also found a
link of $S$ with the total kinetic energy of the system $T$, and a
relationship of Shannon's information entropy in position-space
$S_{r}$ with an experimental quantity i.e. the rms radius of
nuclei and clusters.
In Ref \cite{Lalazissis98} we used another definition of entropy
according to phase-space considerations \cite{Ghosh84}. Thus we
derived an information-theoretic criterion of the quality of a
nuclear density distribution i.e. the larger $S$, the better the
quality of the nuclear model.
Another interesting result \cite{Hall94} is the fact that the
entropy of an N-photon state subjected to Gaussian noise increases
linearly with the logarithm of N.
In Ref. \cite{Panos01a} we considered the single-particle states
of a nucleon in nuclei, a $\Lambda$ in hypernuclei and a valence
electron in atomic clusters. We proposed a connection of $S$ with
the energy $E$ of single-particles states through the relation
\begin{equation}
S=k\ln (\mu E +\nu )
 \label{S-mu}
\end{equation}
where $k$, $\mu$ and $\nu$ depend on the system. It is interesting
that the same relation holds for various systems.

In the present Letter we verify numerically that inequalities
(\ref{Sr-ineq}), (\ref{Sk-ineq}) and (\ref{S-ineq}) hold for a
correlated bosonic system as well, i.e. trapped boson-alkali atoms
\cite{Anderson95,Dalfano99,Fabrocini99} as shown in Table 1. It is
noted that for large $N$ $S_k$ may become negative, but the
important quantity is the net information content $S=S_r+S_k$ of
the system which is positive. We employed density distributions
$\rho({\bf r})$ and $n({\bf k})$ for bosons derived by solving
numerically the Gross-Pitaevskii equation of the form
\begin{equation}
\left[-\frac{\hbar^2}{2m} \nabla^2 + \frac{1}{2}m\omega^2 r^2+
N\frac{4\pi \hbar^2 a }{m} \mid \psi({\bf r}) \mid ^2\right]
\psi({\bf r})=\mu \psi({\bf r}) ,
 \label{gross}
\end{equation}
where $N$ is the number of the atoms, $a$ is the scattering length
of the interaction and $\mu$ is the chemical potential
\cite{Dalfano99}.

For a system of non-interacting bosons in an isotropic harmonic
trap the condensate has a Gaussian form of average width $b$
($b=(\hbar/m\omega)^{1/2})$. If the atoms are interacting, the
shape of the condensate can be changed significantly with respect
to the Gaussian. The ground-state properties of the condensate for
weakly interacting atoms are explained quite successfully by the
non-linear equation (\ref{gross}).

We solved numerically this equation for trapped boson-alkali atoms
(${}^{87}$Rb) systems with parameter $b=12180$ \AA $\,$ (angular
frequency $\omega/\pi=77.78$ Hz) and scattering length $a=52.9$
\AA . In this case the effective atomic size is small compared
both to the trap size and to the interatomic distance ensuring the
diluteness of the gas. The density distribution $\rho({\bf r})$,
obtained in this way, and the momentum distribution $n({\bf k})$,
obtained by taking the Fourier transform of the ground-state wave
function  $\psi({\bf r})$, were inserted into equations
(\ref{Sr1}) and (\ref{Sk1}) to find the values of $S_r$, $S_{k}$
and $S=S_{r}+S_{k}$ as functions of the number of bosons $N$. The
results are shown in Fig. 1. The circles correspond to our
calculated values, while the line to our fitted form
 \[
 S=S_{r}+S_{k}= a+b \ln N
  \]
where $a=6.033$ , $b=0.068$ and $5\times 10^2 < N < 10^6$.

It is noted that the validity of (\ref{gross}) is based on the
condition that the s-wave scattering length be much smaller than
the average distance between atoms and that the number of atoms in
the condensate be much larger than one \cite{Dalfano99}.

 It is seen that a similar functional form holds
approximately for $S$ as function of the number of particles $N$
for fermionic and bosonic systems (correlated atoms in a trap).

Landsberg \cite{Landsberg84} defined the order parameter $\Omega$
as
\begin{equation}
 \Omega = 1-\Delta = 1- \frac{S}{S({\rm max})}
 \label{omega}
 \end{equation}
where $S$ is the information entropy (actual) of the system and
$S({\rm max})$ the maximum entropy accessible to the system. Thus
the concepts of entropy and disorder are decoupled and it is
possible for the entropy and order to increase simultaneously. It
is noted that $\Omega =1$ corresponds to perfect order and
predictability, while $\Omega =0$ means complete disorder and
randomness.

Our results in the present work for $S$ and $S({\rm max})$ allows
us to calculate $\Omega$ as function of the number of particles
$N$ in a system of trapped correlated boson-alkali atoms. The
dependence $\Omega(N)$ is presented in Figure 2. It is seen that
$\Omega$ is an increasing function of $N$. A similar trend has
been observed in Fig. 1 of Ref. \cite{Panos01b}, where $\Omega(N)$
was calculated for nucleons in nuclei and valence electrons in
atomic clusters. As stated in \cite{Panos01b}, our result is in a
way counter-intuitive and indicates that as particles are added in
a correlated quantum-mechanical system, the system becomes more
ordered. The authors in \cite{Landsberg98} studied disorder and
complexity in an ideal Fermi gas of electrons. They observed that
for a small number of electrons the order parameter $\Omega$ is
small, while $\Omega$ increases as one pumps electrons into the
system and the energy levels fill up.

Concluding, we may state that fermionic and correlated bosonic
systems show certain similarities from an information-theoretic
point of view. The information entropy $S$ obeys for both the same
functional form $S=a+b \ln N$. We have also shown that the same
rigorous inequalities (upper and lower limits) hold for both
systems. Finally, we have shown that Landsberg's order parameter
$\Omega$ is an increasing function of the number of particles $N$
for both systems. It is conjectured that the above properties are
universal for fermionic and bosonic many-body quantum systems. It
is also remarkable that those properties hold for systems of
different sizes i.e. ranging from the order of fermis ($10^{-13}$
cm) in nuclei to $10^4$ \AA  $\;(10^{-4}$ cm ) for bosonic
systems.


\bigskip\bigskip
\begin{table}[h]
\caption{Values of $S_r$, $S_k$ and $S$ versus the number of
particles $N$ for a bosonic system (see inequalities (7), (8) and
(9))}. \label{table-1}
\begin{tabular}{ r c c c c c c c c c} \hline

  $N$ & $S_r(min)$& $S_r\,\,\,$& $S_r(max)$& $S_k(min)$& $S_k\,\,\,$
& $S_k(max)$& $S(min)$ &$S\,\,\,\,$& $S(max)$ \\
\hline
 $5\times 10^2$&
 3.797 &3.834 &3.845 &2.590 &2.630 &2.637 &6.434 &6.465 &  6.482\\
&     &           &         &       &       &       &     & \\
 $10^3$&
4.027 &4.100 &4.120 & 2.314 & 2.394 & 2.408 &6.434 &6.494 &6.528\\
&     &           &         &       &       &       &     & \\
$3\times 10^3$&
4.437 &4.599 &4.640 & 1.794 & 1.963 & 1.997 &6.434 &6.562 &6.637 \\
&     &           &         &       &       &       &     & \\
$5\times 10^3$&
4.641 &4.855 &4.907 & 1.527 & 1.746 & 1.794 &6.434 &6.601 &6.701 \\
&     &           &         &       &       &       &     & \\
$7\times 10^3$&
4.778 &5.029 &5.090 & 1.345 & 1.598 & 1.657 &6.434 &6.627 &6.746 \\
&     &           &         &       &       &       &     & \\
$10^4$&
4.925 &5.219 &5.287 & 1.148 & 1.437 & 1.509 &6.434 &6.655 &6.796 \\
&     &           &         &       &       &       &     & \\
$5\times 10^4$&
5.615 &6.113 &6.211 & 0.223 & 0.667 & 0.819 &6.434 &6.780 &7.030 \\
&     &           &         &       &       &       &     & \\
$10^5$&
5.922 &6.511 &6.619 &-0.185 & 0.317 & 0.512 &6.434 &6.828 &7.132\\
&     &           &         &       &       &       &     & \\
$5\times 10^5$&
 6.654 &7.452 &7.577 &-1.142 &-0.533 &-0.220 &6.434 &6.919 &7.357 \\
&     &           &         &       &       &       &     & \\
$10^6$&
6.993 &7.864 &7.992 &-1.557 &-0.920 &-0.560 &6.434 &6.943 &7.432 \\
\hline
\end{tabular}
\end{table}

\bigskip

\begin{figure}
\begin{center}
\begin{tabular}{cc}
\includegraphics[width=6.cm]{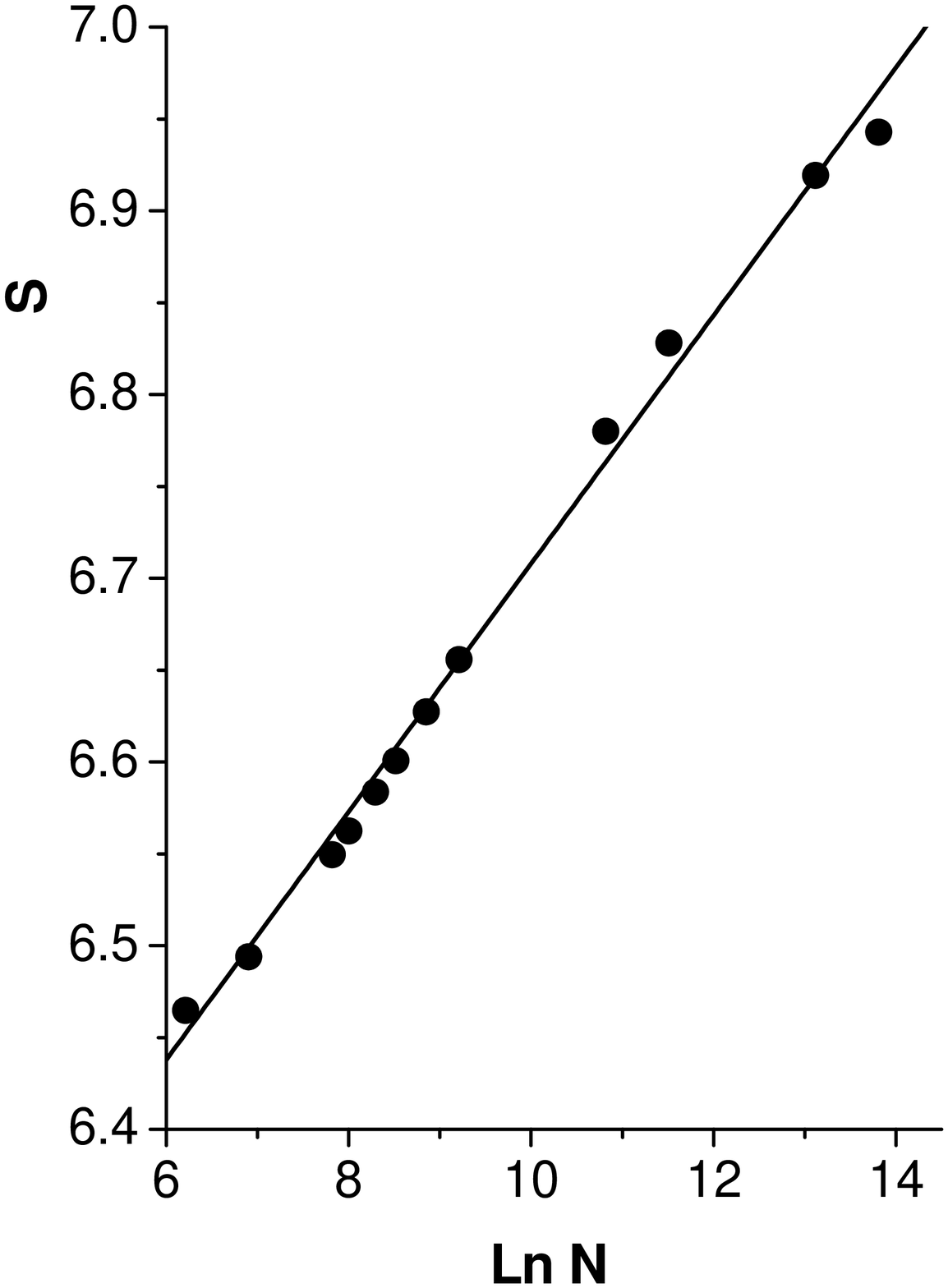} &
\includegraphics[width=6.cm]{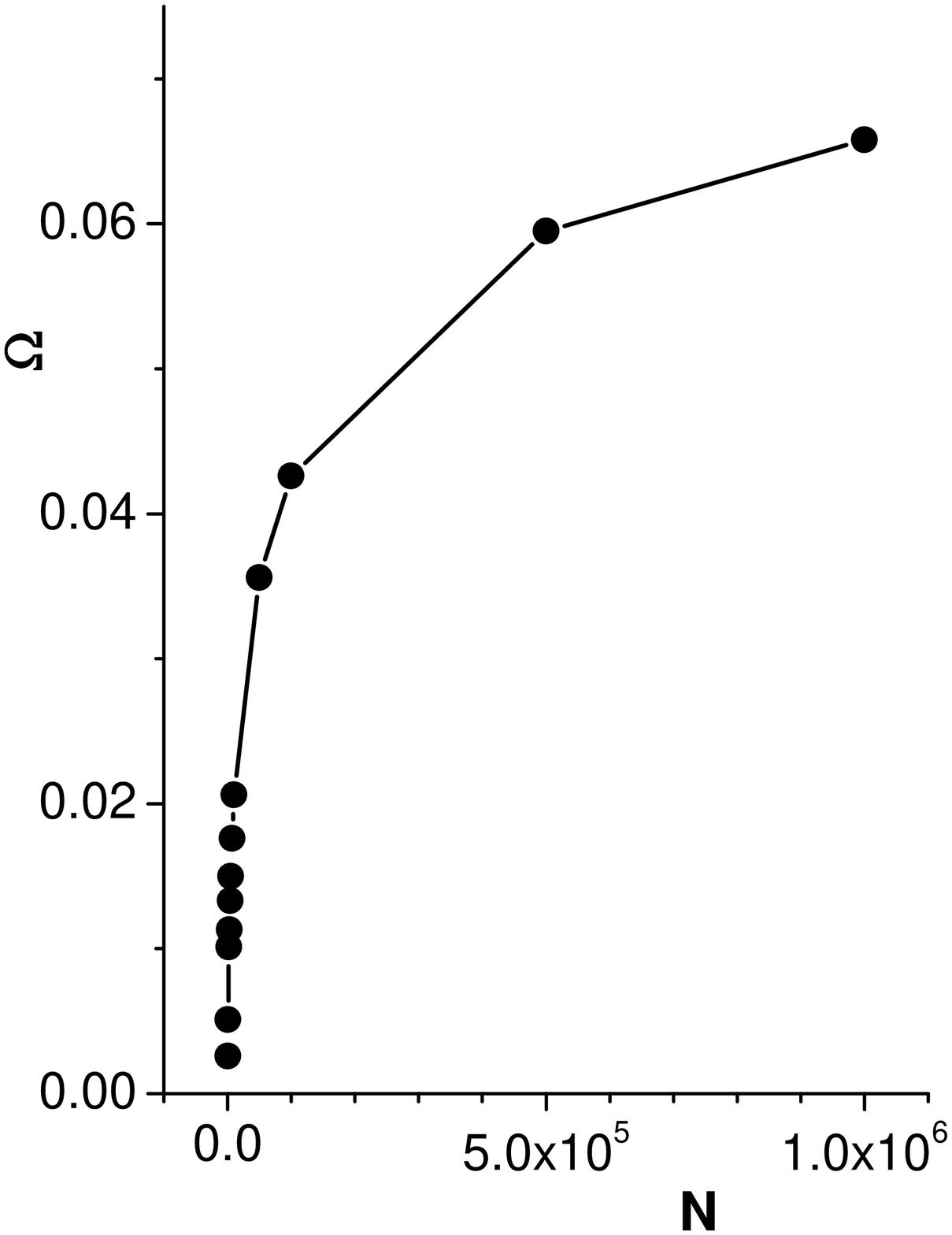}
\end{tabular}
\end{center}
\hspace*{1cm}\begin{minipage}{7cm} \caption{Plot of information
entropy $S=S_r+S_k$ as a function of the number of particles $N$
for a bosonic system.}
\end{minipage}
\hfill
 \vspace*{-1.8cm}
 \hspace*{9cm}\begin{minipage}{7cm}
 \caption{The order parameter $\Omega$ as a function of the
number of particles $N$ for a bosonic system.}
\end{minipage}
\end{figure}

\end{document}